\documentclass[pra, aps, english, twocolumn, hyperref]{revtex4}

\usepackage{graphicx}

\usepackage{amssymb, amsmath, color, rotating}
\begin{document}

\title{Landau damping in a collisionless dipolar Bose gas}

\author{Stefan S. Natu}

\email{snatu@umd.edu}

\affiliation{Condensed Matter Theory Center,  National Institute of Standards and Technology and Joint Quantum Institute, Department of Physics, University of Maryland, College Park, Maryland 20742-4111 USA}

\author{Ryan M. Wilson} 

\affiliation{Joint Quantum Institute, National Institute of Standards and Technology and University of Maryland, College Park, Maryland 20742-4111 USA}

\begin{abstract}
We present a theory for the Landau damping of low energy quasi-particles in a collisionless, quasi-$2$D dipolar Bose gas and produce expressions for the damping rate in uniform and non-uniform systems. Using simple energy-momentum conservation arguments, we show that in the homogeneous system, the nature of the low energy dispersion in a dipolar Bose gas severely inhibits Landau damping of long wave-length excitations. For a gas with contact and dipolar interactions, the damping rate for phonons tends to \textit{decrease} with increasing dipolar interactions; for strong dipole-dipole interactions, phonons are virtually undamped over a broad range of temperature. The damping rate for maxon-roton excitations is found to be significantly larger than the damping rate for phonons.
\end{abstract}
\maketitle
\section{Introduction}
The trapping and cooling of highly magnetic atoms such as $^{52}$Cr, $^{162}$Dy, $^{164}$Dy and $^{168}$Er \cite{pfau2, dysprosium2, dysprosium, erbium} has allowed experimentalists to create Bose-Einstein condensates with strong dipole-dipole interactions \cite{lahayereview}. The long range character of these interactions  leads to phenomena such as geometry dependent mechanical stability \cite{pfau}, anisotropic superfluid velocity \cite{wilson}, novel topological defects and vortex lattice configurations upon rotation \cite{abad, cooper}. A key ingredient responsible for some of the new physics of dipolar Bose gases is the low energy excitation spectrum, which has a phonon-roton character in reduced dimensions \cite{santos, giovanazzi, mazzanti}, resembling that of superfluid He--$4$ \cite{marisreview, mills, griffinbook}. Here we present a theory to describe the damping of these excitations at finite temperature, where the damping is provided by interactions between low lying excitations, mediated by the underlying Bose condensate.

The dynamics of Bose condensed gases with short-range interactions is well studied \cite{Beliaev, szepfalusy, shlyapnikov, fedichev, chung, schieve, liu, stringari, giorgini, giorgini2, griffin, shenoy, stoof} at zero and finite temperature, in homogeneous as well as trapped  gases. In the homogeneous case, there is a continuum of excitations characterized by the familiar Bogoliubov dispersion, which contains linearly dispersing phonons at energies smaller than the chemical potential $\mu$, and free-particle-like excitations at energies larger than $\mu$ \cite{pethick}. At low temperatures and low densities, such as those realized in cold atomic gases, the mean free path of the excitations is comparable to the system size and the system enters the ``collisionless" regime, where the characteristic excitation frequency is much larger than the typical relaxation rate due to collisions $\omega\tau \gg 1$. In this limit, interactions between quasi-particles are primarily mediated by the Bose condensate. A given excitation can decay into two excitations with lower energy, a process known as Beliaev damping \cite{Beliaev}, which dominates at low temperatures $k_{B}T \ll \mu$ \cite{popov}. At higher temperatures, an excitation can also decay by near resonant coupling to transitions between other elementary excitations, a mechanism known as Landau damping. In the context of the dilute Bose gas, Landau damping was first discussed in the low temperature limit by Hohenberg and Martin \cite{martin}, and at higher temperatures by Sz\'epfalusy and Kondor \cite{szepfalusy}. A systematic theory of Landau damping in a dilute Bose gas covering both temperature regimes was later provided by Pitaevskii and Stringari \cite{stringari}. In the opposite limit, where the typical collision time is small $\omega\tau \ll 1$, the damping of collective excitations is governed by two-fluid hydrodynamics \cite{griffin3}. 

In trapped systems, the spectrum of low lying modes is discrete, and the frequencies obtained experimentally match theoretical predictions based on superfluid hydrodynamics \cite{stringari1, jin, ketterle2, ketterle}. At higher energies and temperatures, the dynamics are collision dominated, and the temperature dependence of the excitation frequencies and the associated damping rates has been studied in Refs.~\cite{griffin, shenoy, jin2, fedichev}. At lower energies, the discretization of the collective modes forbids Beliaev damping altogether, and the temperature dependence of the  damping rates observed experimentally \cite{jin2, ketterle2} has been found to be consistent with Landau damping \cite{shlyapnikov, schieve, liu, giorgini2}.

In recent years, the focus has shifted to understanding the properties of Bose condensed gases with long range interactions. At low temperatures, the dynamics of such a condensate is well described by a non-local Gross Pitaevskii equation, which was successfully employed in describing the anisotropic collapse of a dipolar $^{52}$Cr condensate \cite{pfau}. The spectrum of excitations in the dipolar gas has been established in homogeneous and harmonically confined systems \cite{santos, giovanazzi, mazzanti, fischerdipole, wilson2, blakie}. 

In this work, we study the Landau damping of excitations in a quasi-$2$D dipolar Bose gas. We follow the time-dependent mean-field formalism developed by Giorgini \cite{giorgini, giorgini2}, which describes the coupled dynamics of the condensed and non-condensed clouds interacting with short range interactions. Generalizing this formalism to include long range interactions, we obtain the Landau damping rate in a dipolar gas. We remark that the zero temperature version of this theory yields the Beliaev damping rate, which is studied in detail in Refs.~\cite{natu1, ryan1}. 


We find that the Landau damping of low energy, long wave-length excitations is \textit{exponentially} suppressed at low temperatures. This is in departure from the $T^{2}$ dependence of the damping rate expected in $2$D systems with short-range interactions \cite{chung}. Surprisingly we find that the phonon damping rate \textit{decreases} with increasing dipolar interactions, and for sufficiently large dipolar interactions, phonons are virtually undamped at finite temperature. By contrast,  the damping rate for roton-maxon excitations is significantly higher than the phonon damping rate.

This paper is organized as follows: In Sec II we describe the time-dependent mean-field scheme used to compute the damping rates, and obtain formulas for the Landau damping rate in homogeneous and trapped systems in the presence of long range interactions. In Sec III, we specialize to the case of a homogeneous, quasi-$2$D dipolar gas, and compute the Landau damping rate for phonons at low and high temperatures.  To highlight the role of the low energy dispersion in determining the damping rates, we compare the phonon damping rate for a purely dipolar gas with that of a gas with purely contact interactions. We then consider the case where both dipolar and contact interactions are present. In Sec IV we numerically calculate the Landau damping rate at intermediate momenta, where the low energy dispersion relation develops a maxon-roton feature. We discuss the experimental implications of our work in Section V, and summarize our results in Section VI.

\section{Formalism}

The Hamiltonian for a Bose gas interacting with a potential $V_{\text{tot}}(\textbf{r}-\textbf{r}^{'})$ is given by:
\begin{eqnarray}\label{Ham}
{\cal{H}} = \int d\textbf{r}~\Psi^{\dagger}(\textbf{r}, t)\Big[-\frac{\hbar^{2}\nabla^{2}}{2m} - \mu(\textbf{r}) \Big]\Psi(\textbf{r}, t) + \\\nonumber \frac{1}{2}\int d\textbf{r}d\textbf{r}^{'}V_{\text{tot}}(\textbf{r}-\textbf{r}^{'})\Psi^{\dagger}(\textbf{r}, t)\Psi^{\dagger}(\textbf{r}^{'}, t)\Psi(\textbf{r}^{'}, t)\Psi(\textbf{r}, t)
\end{eqnarray}
where $\Psi(\textbf{r}, t)$ is the bosonic annihilation operator at position $\textbf{r}$ and time $t$, $m$ is the mass, and $\mu(\textbf{r})$ is the spatially varying chemical potential. In the presence of a trapping potential $U(\textbf{r})$, $\mu(\textbf{r}) = \mu_{0} - U(\textbf{r})$, where $\mu_{0}$ is the chemical potential in the center of the trap. The interaction potential is assumed to be a sum of two terms $V_{\text{tot}}(\textbf{r} -\textbf{r}^{'}) = g\delta(\textbf{r}-\textbf{r}^{'}) + V_{\text{lr}}(\textbf{r}-\textbf{r}^{'})$ where $g = \frac{4\pi \hbar^{2}a}{m}$ parametrizes the contact (short-range) part of the potential with a $3$D s-wave scattering length $a$, and all the long range terms are included in $V_{\text{lr}}(\textbf{r}-\textbf{r}^{'})$. Throughout we will assume that the potential satisfies $V_{\text{tot}}(\textbf{r}) = V_{\text{tot}}(-\textbf{r})$. For dipolar bosons with dipole moments $d$, that are oriented along the $z$ axis \cite{lahayereview}:
\begin{equation}\label{vdip}
V_{\text{lr}}(\textbf{r}-\textbf{r}^{'}) = d^{2}\frac{1 - 3\cos^{2}(\theta)}{|\textbf{r} -\textbf{r}^{'}|^{3}}
\end{equation}
where $\theta$ is the angle between the vector $\textbf{r} -\textbf{r}^{'}$ and the $z$ axis. 

The equation of motion for $\Psi$ is obtained using the Heisenberg prescription:
\begin{eqnarray}\label{psieom}
i\hbar\partial_{t}\Psi(\textbf{r}, t) = \Biggl[\Big(-\frac{\hbar^{2}\nabla^{2}}{2m} - \mu(\textbf{r})\Big) + \int d\textbf{r}^{'} V_{\text{tot}}(\textbf{r}-\textbf{r}^{'})\times\\\nonumber|\Psi(\textbf{r}^{'}, t)|^{2}\Biggr]\Psi(\textbf{r}, t).
\end{eqnarray}
At temperatures below the Bose-condensation temperature, the bosonic annihilation operator can be decomposed as a sum of two parts: $\Psi(\textbf{r}, t) = \phi(\textbf{r}, t) + \psi(\textbf{r}, t)$ where $\phi(\textbf{r}, t) = \langle \Psi(\textbf{r}, t) \rangle$  is a complex order parameter representing the condensate, and $\psi(\textbf{r}, t)$ represents the non-condensed atoms, which by definition have the property $\langle \psi(\textbf{r}, t) \rangle = 0$. The averages here in general denote non-equilibrium averages, as we allow for fluctuations of the condensed and non-condensed atoms. Equilibrium averages will be denoted by $\langle ..\rangle_{0}$. 

Decomposing the term proportional to $\Psi^{\dagger}\Psi^{\dagger}\Psi$ in Eq.~\ref{psieom} into condensate and non-condensate contributions, the equation of motion for the condensate field reads:
\begin{eqnarray}\label{phieom}
i\hbar\partial_{t}\phi(\textbf{r}, t) = -\Big(\frac{\hbar^{2}\nabla^{2}}{2m}+\mu(\textbf{r})\Big)\phi(\textbf{r}, t)+ \int d\textbf{r}^{'}V_{\text{tot}}(\textbf{r}-\textbf{r}^{'})\\\nonumber\times\Big[|\phi(\textbf{r}^{'}, t)|^{2}\phi(\textbf{r}, t) +\phi(\textbf{r}^{'}, t)n(\textbf{r}^{'},\textbf{r}, t)+ \\\nonumber \phi^{*}(\textbf{r}^{'}, t)m(\textbf{r}^{'}, \textbf{r}, t) + n(\textbf{r}^{'}, t)\phi(\textbf{r}, t)\Big]
\end{eqnarray}
where we have introduced normal and anomalous densities 
\begin{eqnarray}\label{nadens}
n(\textbf{r}^{'}, \textbf{r}, t) = \langle \psi^{\dagger}(\textbf{r}^{'}, t)\psi(\textbf{r}, t) \rangle\\\nonumber
m(\textbf{r}^{'}, \textbf{r}, t) = \langle \psi(\textbf{r}^{'}, t)\psi(\textbf{r}, t) \rangle 
\end{eqnarray}
and we use the short-hand notation $n(\textbf{r}, t) = \langle \psi^{\dagger}(\textbf{r}, t)\psi(\textbf{r}, t) \rangle$ and $m(\textbf{r}, t) = \langle \psi(\textbf{r}, t)\psi(\textbf{r}, t) \rangle$. Setting $n$ and $m$ to zero in Eq.~\ref{phieom}, one obtains a non-local Gross-Pitaevskii equation (GPE), which describes the dynamics of a Bose condensate at zero temperature interacting with long range interactions \cite{lahayereview, pfau}. 

In general we are interested in small, time-dependent fluctuations of the condensate about its equilibrium value, which is obtained by solving the GPE in equilibrium. To describe the fluctuations, we expand $\phi(\textbf{r}, t) = \phi_{0}(\textbf{r}) + \delta \phi(\textbf{r}, t)$, where $\phi_{0}(\textbf{r})$ is the equilibrium condensate density, which we assume without any loss of generality to be real \cite{caveat}, and $\delta\phi(\textbf{r}, t)$ corresponds to small, time-dependent fluctuations about the stationary state. At finite temperature, the fluctuations of the condensate  are coupled to fluctuations of the non-condensate densities, which can be written in the form: $n(\textbf{r}, \textbf{r}^{'}, t) = n_{\text{eq}}(\textbf{r}, \textbf{r}^{'}) + \delta n(\textbf{r}, \textbf{r}^{'}, t)$ and $m(\textbf{r}, \textbf{r}^{'}, t) = \delta m(\textbf{r}, \textbf{r}^{'}, t)$. Here we assume that the anomalous density of the thermal cloud is zero in equilibrium ($m_{\text{eq}}(\textbf{r}, t) = \langle \psi(\textbf{r}, t)\psi(\textbf{r}, t)  \rangle_{0} = 0$), an approximation due to Popov \cite{popov}. As discussed by Giorgini \cite{giorgini, giorgini2}, this approximation yields an inadequate result for the \textit{real} part of the phonon frequency shift, but leaves the imaginary part (which contributes to damping) unchanged. A detailed description of the Popov approximation and its implications can be found in Ref.~\cite{griffin3}.

Linearizing Eq.~\ref{phieom} in the Popov approximation, the equation of motion for the fluctuations of the condensate reads:
\begin{eqnarray}\label{condfluc}
i\partial_{t}\delta\phi(\textbf{r}, t) = {\cal{\hat L}}_{0}~\delta\phi(\textbf{r}, t) + \int d\textbf{r}^{'}V_{\text{tot}}(\textbf{r}-\textbf{r}^{'})\times\\\nonumber
\Big[\phi_{0}(\textbf{r})\phi_{0}(\textbf{r}^{'})(\delta \phi^{*}(\textbf{r}^{'}, t) + \delta \phi(\textbf{r}^{'}, t))+\\\nonumber n_{0}(\textbf{r}^{'}, \textbf{r})\delta\phi(\textbf{r}^{'}, t) + \phi_{0}(\textbf{r}^{'})(\delta n(\textbf{r}^{'}, \textbf{r}, t) + \\\nonumber \delta m(\textbf{r}^{'}, \textbf{r}, t)) + \phi_{0}(\textbf{r})\delta n(\textbf{r}^{'}, t)\Big]
\end{eqnarray}
where we define the operator
\begin{equation}\label{ldef}
{\cal{\hat L}}_{0} \equiv -\frac{\hbar^{2}\nabla^{2}}{2m} - \mu(\textbf{r}) + \int d\textbf{r}^{'}V_{\text{tot}}(\textbf{r}-\textbf{r}^{'})n_{\text{tot}}(\textbf{r}^{'})
\end{equation}
and $n_{\text{tot}}(\textbf{r}) = |\phi_{0}(\textbf{r})|^{2} + n_{\text{eq}}(\textbf{r})$ is the total density. In general we define $n_{\text{tot}}(\textbf{r}^{'},\textbf{r}) = \phi_{0}(\textbf{r}^{'})\phi_{0}(\textbf{r}) + n_{\text{eq}}(\textbf{r}^{'},\textbf{r})$.

The corresponding Hamiltonian for the non-condensed atoms can be decomposed into two parts:
\begin{equation}\label{hnonc}
{\cal{H}}_{\text{nc}} = {\cal{H}}_{0} + {\cal{H}}_{\delta}
\end{equation}
where 
\begin{eqnarray}\label{bogpiece}
{\cal{H}}_{0} = \int d\textbf{r}d\textbf{r}^{'} \Big\{\psi^{\dagger}(\textbf{r}, t){\cal{\hat L}}_{0}\psi(\textbf{r}, t) + V_{\text{tot}}(\textbf{r} -\textbf{r}^{'})n_{\text{tot}}(\textbf{r}^{'}, \textbf{r})\hspace{2mm}\\\nonumber \times \Big[\psi^{\dagger}(\textbf{r}^{'}, t)\psi(\textbf{r}, t) +\psi^{\dagger}(\textbf{r},t)\psi(\textbf{r}^{'},t)\Big] +  \\\nonumber \phi_{0}(\textbf{r})\phi_{0}(\textbf{r}^{'})\times \Big[\psi^{\dagger}(\textbf{r}^{'}, t)\psi^{\dagger}(\textbf{r}, t)+ \psi(\textbf{r}, t)\psi(\textbf{r}^{'}, t)\Big]\Big\}
\end{eqnarray}
in Popov approximation. The Hamiltonian ${\cal{H}}_{0}$ couples the non-condensed atoms to the equilibrium condensate and non-condensate densities, and does not contain any condensate fluctuations. 

The latter are described by the Hamiltonian ${\cal{H}}_{\delta}$, containing terms linear in $\delta\phi$:
 
\begin{eqnarray}\label{flucpiece}
{\cal{H}}_{\delta} = \frac{1}{2}\int d\textbf{r}d\textbf{r}^{'} V_{\text{tot}}(\textbf{r}-\textbf{r}^{'})\Big[\phi_{0}(\textbf{r})\Big\{(\delta\phi(\textbf{r}, t)+ \\\nonumber \delta\phi^{*}(\textbf{r}, t))\times\psi^{\dagger}(\textbf{r}^{'}, t)\psi(\textbf{r}^{'}, t) +\delta\phi(\textbf{r}^{'}, t)\times \\\nonumber\psi^{\dagger}(\textbf{r}^{'}, t)\psi(\textbf{r}, t)+  \delta\phi^{*}(\textbf{r}^{'}, t)\psi^{\dagger}(\textbf{r},t)\psi(\textbf{r}^{'},t)+ \\\nonumber \delta\phi(\textbf{r}^{'},t)\psi^{\dagger}(\textbf{r},t)\psi^{\dagger}(\textbf{r}^{'},t) + \delta\phi^{*}(\textbf{r}^{'},t)\times\\\nonumber \psi(\textbf{r}^{'},t)\psi(\textbf{r},t)\Big\} + \textbf{r}\leftrightarrow \textbf{r}^{'}\Big]
\end{eqnarray}
where $\textbf{r} \rightarrow \textbf{r}^{'}$ above refers to interchanging $\textbf{r}$ and $\textbf{r}^{'}$ for every term in the preceding curly brackets. In the weakly interacting gas, the condensate density greatly exceeds the density of the non-condensed atoms ($|\phi_{0}|^{2} \gg n_{\text{eq}}$). As a result, interactions between the non-condensed atoms are mainly mediated by the condensate, and therefore we have ignored terms involving only non-condensed atoms (all terms in ${\cal{H}}_{\delta}$ containing four non-condensed field operators).


The Hamiltonian ${\cal{H}}_{0}$ given by Eq.~\ref{bogpiece} can be diagonalized via the usual Bogoliubov transformation:
\begin{eqnarray}\label{bogdef}
\psi(\textbf{r}, t) = \sum_{i}u_{i}(\textbf{r})a_{i}(t) +v^{*}_{i}(\textbf{r})a^{\dagger}_{i}(t) \\\nonumber
\psi^{\dagger}(\textbf{r}, t) = \sum_{i}u^{*}_{i}(\textbf{r})a^{\dagger}_{i}(t) +v_{i}(\textbf{r})a_{i}(t)
\end{eqnarray}
where $a_{i}(t)$ denotes the bosonic quasi-particle annihilation operator at time $t$, and satisfies the usual bosonic commutation relations \cite{pethick}. The complex functions $u_{i}$ and $v_{i}$ obey $\int d\textbf{r} d\textbf{r}^{'}(u^{*}_{i}(\textbf{r})u_{j}(\textbf{r}^{'}) - v^{*}_{i}(\textbf{r})v_{j}(\textbf{r}^{'})) = \delta_{ij}\delta(\textbf{r}-\textbf{r}^{'})$ where $\delta_{ij}$ is the Kronecker delta symbol. 

The resulting Hamiltonian describes a non-interacting gas of Bogoliubov quasi-particles:
\begin{equation}\label{onlybogham}
{\cal{H}}_{0} = \sum_{i}E_{i}a^{\dagger}_{i}a_{i}
\end{equation}
where $E_{i}$ are the quasi-particle energies obtained by solving the Bogoliubov equations:

\begin{eqnarray}\label{edef}
E_{i}u_{i}(\textbf{r}) = {\cal{\hat L}}[u_{i}(\textbf{r})] + {\cal{\hat{M}}}[v_{i}(\textbf{r})] \\\nonumber
-E_{i}v_{i}(\textbf{r}) = {\cal{\hat L}}[v_{i}(\textbf{r})] + {\cal{\hat{M}}}[u_{i}(\textbf{r})]
\end{eqnarray}
where we define 
\begin{eqnarray}\label{lmdef}
{\cal{\hat L}}[f(\textbf{r})] = {\cal{\hat L}}_{0}f(\textbf{r}) + \int d\textbf{r}^{'}V_{\text{tot}}(\textbf{r}-\textbf{r}^{'})n_{\text{tot}}(\textbf{r}^{'}, \textbf{r})f(\textbf{r}^{'}) \hspace{2mm}\\\nonumber
{\cal{\hat M}}[f(\textbf{r})] = \int d\textbf{r}^{'}V_{\text{tot}}(\textbf{r}-\textbf{r}^{'})\phi_{0}(\textbf{r})\phi_{0}(\textbf{r}^{'})f(\textbf{r}^{'}). \hspace{12mm}
\end{eqnarray}

A key difference between the Bogoliubov equations above and those for a gas interacting with purely contact interactions arises in the non-local term in the definition of ${\cal{\hat L}}[f(\textbf{r})]$ and ${\cal{\hat M}}[f(\textbf{r})]$. This term represents the ``exchange" (Fock) contribution to the interaction. For a gas with short-range interactions, this term is identical to the ``direct" (Hartree) contribution, and together they give rise to a factor of $2$ in the interaction term. In the presence of  long range interactions, these contributions must be treated separately, and the problem of solving the Bogoliubov equations becomes computationally intensive \cite{ronen, blakie2, ticknor}.

The equation for the fluctuations of the condensate Eq.~\ref{condfluc} can be expressed in terms of the Bogoliubov operators $a_{i}$ and $a^{\dagger}_{i}$ by expanding the normal and anomalous densities as follows:
\begin{eqnarray}\label{ndensbog}
n(\textbf{r}^{'}, \textbf{r}, t) = \sum_{ij}u^{*}_{i}(\textbf{r}^{'})u_{j}(\textbf{r})f_{ij} + v_{i}(\textbf{r}^{'})v^{*}_{j}(\textbf{r})f_{ji} + \\\nonumber u^{*}_{i}(\textbf{r}^{'})v^{*}_{j}(\textbf{r})g^{*}_{ij} +v_{i}(\textbf{r}^{'})u_{j}(\textbf{r})g_{ij} + v_{i}(\textbf{r}^{'})v^{*}_{j}(\textbf{r})
\end{eqnarray}
and
\begin{eqnarray}\label{adensbog}
m(\textbf{r}^{'}, \textbf{r}, t) = \sum_{ij}u_{i}(\textbf{r}^{'})v^{*}_{j}(\textbf{r})f_{ji} + v^{*}_{i}(\textbf{r}^{'})u_{j}(\textbf{r})f_{ij} + \\\nonumber u_{i}(\textbf{r}^{'})u_{j}(\textbf{r})g_{ij} +v^{*}_{i}(\textbf{r}^{'})v^{*}_{j}(\textbf{r})g^{*}_{ij} + u_{i}(\textbf{r}^{'})v^{*}_{j}(\textbf{r}),
\end{eqnarray}
where it is convenient to define the functions:
\begin{eqnarray}\label{bogdensdef}
f_{ij}(t) \equiv \langle a^{\dagger}_{i}(t)a_{j}(t) \rangle\\\nonumber
g_{ij}(t) \equiv \langle a_{i}(t)a_{j}(t) \rangle.
\end{eqnarray}

The equation for the fluctuations of the condensate (Eq.~\ref{condfluc}) is thus coupled to the dynamics of the non-condensed atoms via $f_{ij}$ and $g_{ij}$. Physically, $f_{ij}$ corresponds to the creation and annihilation of a quasi-particle with energy $E_{i}$ and $E_{j}$, and contributes to Landau damping, while $g_{ij} (g^{*}_{ij})$ corresponds to the annihilation (creation) of two quasi-particles with energies $E_{i}$ and $E_{j}$, and contributes to Beliaev damping \cite{giorgini}. Henceforth we focus on the dynamics of $f_{ij}$ alone.  

The equation for the fluctuations of the condensate (Eq.~\ref{condfluc}) can be written in terms of $f_{ij}(t)$ as:
\begin{widetext}
\begin{eqnarray}\label{condfs}
i\hbar\partial_{t}\delta\phi(\textbf{r}, t) = {\cal{\hat L}}[\delta\phi(\textbf{r}, t)] + {\cal{\hat M}}[\delta \phi^{*}(\textbf{r})] + \sum_{ij}f_{ij}(t) \int d\textbf{r}^{'}V_{\text{tot}}(\textbf{r}-\textbf{r}^{'})\Big(\phi_{0}(\textbf{r}^{'})\Big[u^{*}_{i}(\textbf{r}^{'})u_{j}(\textbf{r}) + v_{j}(\textbf{r}^{'})v^{*}_{i}(\textbf{r})+ \\\nonumber v^{*}_{i}(\textbf{r})u_{j}(\textbf{r}^{'}) + v^{*}_{i}(\textbf{r}^{'})u_{j}(\textbf{r})\Big]+ \phi_{0}(\textbf{r})\Big[u^{*}_{i}(\textbf{r}^{'})u_{j}(\textbf{r}^{'}) + v_{i}(\textbf{r}^{'})v^{*}_{j}(\textbf{r}^{'})\Big]\Big) + ...
\end{eqnarray}
\end{widetext}
The $...$ at the end of Eq.~\ref{condfs} refers to the terms proportional to $g_{ij}$ that give rise to Beliaev damping, which we do not study here. 

Using Heisenberg's equation of motion, $i\hbar \partial_{t}\langle \hat O \rangle = \langle [O, {\cal{H}}_{\text{nc}}]\rangle$, the equation of motion for $f_{ij}$ reads:
\begin{widetext}
\begin{eqnarray}\label{feq}
i\hbar\partial_{t}f_{ij}(t) = (\epsilon_{j}-\epsilon_{i})f_{ij}(t) + \frac{(n^{\text{th}}_{i}-n^{\text{th}}_{j})}{2}\int d\textbf{r}d\textbf{r}^{'}V_{\text{tot}}(\textbf{r}-\textbf{r}^{'})\Big[2\phi_{0}(\textbf{r})(\delta\phi(\textbf{r}, t)+\delta\phi^{*}(\textbf{r}, t))\times\{u^{*}_{j}(\textbf{r}^{'})u_{i}(\textbf{r}^{'}) +\\\nonumber v^{*}_{i}(\textbf{r}^{'})v_{j}(\textbf{r}^{'})\} + 2(\phi_{0}(\textbf{r})\delta\phi(\textbf{r}^{'}, t)+\phi_{0}(\textbf{r}^{'})\delta\phi^{*}(\textbf{r}, t))\{u^{*}_{j}(\textbf{r}^{'})u_{i}(\textbf{r}) + v_{j}(\textbf{r}^{'})v^{*}_{i}(\textbf{r})\} + (\phi_{0}(\textbf{r})\delta\phi(\textbf{r}^{'}, t) + \phi_{0}(\textbf{r}^{'})\delta\phi(\textbf{r}, t))\times\\\nonumber\{u^{*}_{j}(\textbf{r})v_{i}(\textbf{r}^{'}) + v_{i}(\textbf{r})u^{*}_{j}(\textbf{r}^{'})\} + (\phi_{0}(\textbf{r}) \delta\phi^{*}(\textbf{r}^{'}, t) + \phi_{0}(\textbf{r}^{'}) \delta\phi^{*}(\textbf{r}, t))\{u_{i}(\textbf{r})v^{*}_{j}(\textbf{r}^{'}) + v^{*}_{j}(\textbf{r})u_{i}(\textbf{r}^{'})\} \Big]
\end{eqnarray}
\end{widetext}
where we define the equilibrium density of the quasi-particles at temperature $T$,  $n^{\text{th}}_{i} = \langle a^{\dagger}_{i}a_{i} \rangle_{0} = \frac{1}{e^{\beta E_{i}} -1}$ where $\beta = 1/k_{B}T$. 


Thus the fluctuations of the condensate are coupled to the fluctuations of the non-condensed atoms and vice versa. The full self-consistent solution of this system of equations is highly non-trivial. In order to solve the coupled system of equations  we assume that the condensate fluctuates with frequency $\omega$ as $\delta\phi(\textbf{r}, t) = \delta\phi_{1}(\textbf{r})e^{-i\omega t}$ and $\delta\phi^{*}(\textbf{r}, t) = \delta\phi_{2}(\textbf{r})e^{-i\omega t}$ and take the Fourier transform of Eq.~\ref{feq} to get:
\begin{widetext}
\begin{eqnarray}\label{fourierfeq}
f_{ij}(\omega) = \frac{(n^{\text{th}}_{i}-n^{\text{th}}_{j})}{2(\hbar\omega - (\epsilon_{j}-\epsilon_{i}))}\int d\textbf{r}d\textbf{r}^{'}V_{\text{tot}}(\textbf{r}-\textbf{r}^{'})\Big[2\phi_{0}(\textbf{r})(\delta\phi_{1}(\textbf{r})+\delta\phi_{2}(\textbf{r}))\times\{u^{*}_{j}(\textbf{r}^{'})u_{i}(\textbf{r}^{'}) +\\\nonumber v^{*}_{i}(\textbf{r}^{'})v_{j}(\textbf{r}^{'})\} + 2(\phi_{0}(\textbf{r})\delta\phi_{1}(\textbf{r}^{'})+\phi_{0}(\textbf{r}^{'})\delta\phi_{2}(\textbf{r}))\{u^{*}_{j}(\textbf{r}^{'})u_{i}(\textbf{r}) + v_{j}(\textbf{r}^{'})v^{*}_{i}(\textbf{r})\} + (\phi_{0}(\textbf{r})\delta\phi_{1}(\textbf{r}^{'}) + \phi_{0}(\textbf{r}^{'})\delta\phi_{1}(\textbf{r}))\times\\\nonumber\{u^{*}_{j}(\textbf{r})v_{i}(\textbf{r}^{'}) + v_{i}(\textbf{r})u^{*}_{j}(\textbf{r}^{'})\} + (\phi_{0}(\textbf{r}) \delta\phi_{2}(\textbf{r}^{'}) + \phi_{0}(\textbf{r}^{'}) \delta\phi_{2}(\textbf{r}))\{u_{i}(\textbf{r})v^{*}_{j}(\textbf{r}^{'}) + v^{*}_{j}(\textbf{r})u_{i}(\textbf{r}^{'})\} \Big].
\end{eqnarray}
\end{widetext}

In the absence of any coupling between the condensed and non-condensed atoms, the fluctuations of the condensate (Eq.~\ref{condfs}) obey the Bogoliubov equations with real frequency $\omega_{0}$:
\begin{eqnarray}
\hbar\omega_{0}\left(\begin{array}{cc} \delta\phi^{0}_{1}(\textbf{r})\\ \delta\phi^{0}_{2}(\textbf{r})\end{array}\right) = \left(\begin{array}{cc} {\cal{\hat L}}& {\cal{\hat M}}\\ -{\cal{\hat M}}^{*}&-{\cal{\hat L}}^{*} \\ \end{array}\right)\left(\begin{array}{cc} \delta\phi^{0}_{1}(\textbf{r}) \\ \delta\phi^{0}_{2}(\textbf{r}) \end{array}\right)
\end{eqnarray}
and satisfy the normalization criterion $\int d\textbf{r}(|\delta\phi^{0}_{1}(\textbf{r})|^{2} -|\delta\phi^{0}_{2}(\textbf{r})|^{2}) = 1$.

In the limit of small non-condensate density, we may treat the condensate-non-condensed atom coupling as a perturbation, and write $\delta\phi_{1,2} = \delta\phi^{0}_{1,2} + \delta\phi^{'}_{1,2}$, where the corrections to the oscillation amplitude are chosen to be orthogonal to those in the absence of condensate-noncondensate coupling: 
\begin{equation}\label{modecrit}
\int d\textbf{r}\Big(\delta\phi^{0}_{1}(\textbf{r})\delta\phi^{*'}_{1}(\textbf{r}) - \delta\phi^{0}_{2}(\textbf{r})\delta\phi^{*'}_{2}(\textbf{r})\Big)= 0.
\end{equation}

Similarly, we write the perturbed eigenfrequency as $\omega  = \omega_{0} + \delta\omega - i\Gamma_{\text{L}}$ where $\delta\omega$ is the correction to the real part of the normal mode eigenfrequency and $\Gamma_{\text{L}}$ denotes the Landau damping coefficient. 

To lowest order, the equation governing the fluctuations of the condensate in the presence of coupling to the non-condensed atoms reads:
\begin{equation}\label{fouriercondfs}
\hbar\omega\delta\phi_{1}(\textbf{r}) = {\cal{\hat L}}[\delta\phi^{0}_{1}(\textbf{r})] + {\cal{\hat M}}[\delta \phi^{0}_{2}(\textbf{r})] + \sum_{ij}f_{ij}(\omega_{0})A_{ij}(\textbf{r})
\end{equation}
where we have replaced all the $\delta\phi$ and $\delta\phi^{*}$ terms in the RHS, including in the expression for $f_{ij}$ (Eq.~\ref{fourierfeq}) with $\delta\phi^{0}_{1}$ and $\delta\phi^{0}_{2}$, and $\omega$ with $\omega_{0}$. We also introduce the shorthand
\begin{eqnarray}\label{aij}
A_{ij}(\textbf{r}) = \int d\textbf{r}^{'}V_{\text{tot}}(\textbf{r}-\textbf{r}^{'})\Big(\phi_{0}(\textbf{r}^{'})\Big[u^{*}_{i}(\textbf{r}^{'})u_{j}(\textbf{r}) + \\\nonumber  v_{j}(\textbf{r}^{'})v^{*}_{i}(\textbf{r})+ u_{i}(\textbf{r}^{'})v^{*}_{j}(\textbf{r}) + u_{i}(\textbf{r})v^{*}_{j}(\textbf{r}^{'}))\Big]+\\\nonumber  \phi_{0}(\textbf{r})\Big[u^{*}_{i}(\textbf{r}^{'})u_{j}(\textbf{r}^{'}) + v_{i}(\textbf{r}^{'})v^{*}_{j}(\textbf{r}^{'})\Big]\Big).
\end{eqnarray}

To obtain the damping rate,  we multiply Eq.~\ref{fouriercondfs} and the corresponding equation for $\delta\phi_{2}$ by  $\delta\phi^{*}_{1}$ and $\delta\phi^{*}_{2}$ respectively, take the difference of the equations and integrate over space to find \cite{giorgini}:
\begin{equation}\label{deltaphi}
\hbar\omega = \hbar\omega_{0} + \sum_{ij}f_{ij}(\omega_{0})\int d\textbf{r}\Big(A_{ij}(\textbf{r})\delta\phi^{*}_{1}(\textbf{r}) + A^{*}_{ji}(\textbf{r})\delta\phi^{*}_{2}(\textbf{r})\Big)
\end{equation}
The damping rate, which is the focus of this work is given by the imaginary part of $\omega$:
\begin{equation}\label{landau}
\Gamma_{\text{L}}(\omega_{0}) = \frac{\pi}{\hbar}\sum_{ij}(n^{\text{th}}_{i}-n^{\text{th}}_{j})\delta(\hbar\omega_{0} - (E_{j}-E_{i}))|{\cal{A}}_{ij}|^{2}
\end{equation}
where we have used Eqs.~\ref{fourierfeq}, \ref{aij} and \ref{deltaphi} to express $f_{ij}$ in terms of $A_{ij}$, and we define the matrix element:

\begin{equation}\label{landau2}
{\cal{A}}_{ij} = \int d\textbf{r}\Big(A_{ij}(\textbf{r})\delta\phi^{*}_{1}(\textbf{r}) + A^{*}_{ji}(\textbf{r})\delta\phi^{*}_{2}(\textbf{r})\Big)
\end{equation}

The formula for the Landau damping rate given by Eq.~\ref{landau} is the main result of this section. We remind the reader that in deriving this result,  we have not assumed any particular form of the two-body interaction potential, only that it is symmetric with respect to the interchange of the two particles. Furthermore, our equations are general and may be applied to study the Landau damping of collective modes in inhomogeneous systems, such as \textit{trapped} atomic gases.

\subsection{Landau damping in a homogeneous gas}

In general, solving for the finite temperature damping rates of a trapped gas is numerically intensive \cite{shlyapnikov, griffin, shenoy, stoof}. In the presence of long range interactions, this is further exacerbated by the exchange interaction, which is non-local in nature. As a result, calculating the Bogoliubov co-efficients in a trapped dipolar gas is extremely challenging, and practical only in certain simple cases \cite{blakie, ronen, ticknor}. 

In the rest of this work, we focus instead on a \textit{homogeneous} dipolar gas. In a homogeneous system, the condensate density $\phi_{0}(\textbf{r}) = \sqrt{n_{0}}$ is  constant throughout space. Furthermore, in order to obtain semi-analytic results, we will neglect the exchange contribution to the Bogoliubov equations of motion arising from the non-condensed atoms, and retain only the condensate contribution. In other words, we approximate $n_{\text{tot}}(\textbf{r}^{'}, \textbf{r}) \approx \phi_{0}(\textbf{r})\phi_{0}(\textbf{r}^{'}) = \phi^{2}_{0}$. We expect this approximation to be quantitatively accurate at temperatures $k_{B}T \leq \mu$, where the condensate is not significantly depleted. In this limit, the homogeneous  Bogoliubov equations of motion can be readily solved in Fourier space, and we will use the resulting solutions in subsequent computations. For sufficiently large systems, our calculation can be readily generalized to include the effects of a trap via a local density approximation \cite{giorgini2}. 

In the homogeneous geometry, the fluctuations of the condensate take the plane-wave form: $\delta\phi(\textbf{r}) = \frac{1}{\sqrt{V}}e^{-i\textbf{k}\cdot\textbf{r}}u_{\text{k}}$ and $\delta\phi^{*}(\textbf{r}) = \frac{1}{\sqrt{V}}e^{-i\textbf{k}\cdot\textbf{r}}v_{\textbf{k}}$;  $u_{\textbf{p}} = \frac{1}{\sqrt{V}}u_{\textbf{p}}e^{-i\textbf{p}\cdot\textbf{r}}$, and $v_{\textbf{p}} = \frac{1}{\sqrt{V}}v_{\textbf{p}}e^{-i\textbf{p}\cdot\textbf{r}}$, where $u_{\textbf{k}}$ and $v_{\textbf{k}}$ are the Bogoliubov coherence factors which read:
\begin{eqnarray}\label{bogqp}
 u_{\textbf{k}} = \sqrt{\frac{1}{2}\Big(\frac{\epsilon_{k} + V_{\text{tot}}(\textbf{k})}{E_{\textbf{k}}} + 1\Big)} \hspace{20mm}\\\nonumber
 v_{\textbf{k}} = -\text{sgn}(V_{\text{tot}}(\textbf{k}))\sqrt{\frac{1}{2}\Big(\frac{\epsilon_{k} + V_{\text{tot}}(\textbf{k})}{E_{\textbf{k}}} - 1\Big)}
\end{eqnarray}
where we denote $V_{\text{tot}}(\textbf{k}) = \int d\textbf{r}e^{-i\textbf{k}\cdot(\textbf{r}-\textbf{r}^{'})}V_{\text{tot}}(\textbf{r}-\textbf{r}^{'})$ as the Fourier transform of the two-body interaction potential. Here $\epsilon_{k} = \hbar^{2}k^{2}/2m$ and the dispersion $E_{\textbf{k}} = \sqrt{\epsilon_k(\epsilon_k + 2V_{\text{tot}}(\textbf{k})n_{0})}$. 

Inserting the expressions for the fluctuations of the condensate and the non-condensed atoms into Eqs.~\ref{aij}, and \ref{landau}, and integrating over space, the formula for the Landau damping rate for a quasi-particle with momentum $\textbf{k}$ and energy $\hbar\omega_{k}$ in a homogeneous system reads:
\begin{eqnarray}\label{landauh}
\Gamma_{\text{L}}(\textbf{k}) = -\frac{\pi n_{0}}{\hbar}\sum_{\textbf{p}\textbf{q}}(n^{\text{th}}_{\textbf{q}}-n^{\text{th}}_{\textbf{p}})~\delta(\hbar\omega_{\textbf{k}} - E_{\textbf{q}} + E_{\textbf{p}})\times\\\nonumber{\cal{A}}^{\textbf{k}}_{\textbf{p}\textbf{q}}{\cal{A}}^{\textbf{k}}_{\textbf{q}\textbf{p}}
\end{eqnarray}
where
\begin{widetext}
\begin{eqnarray}\label{aijh}
{\cal{A}}^{\textbf{k}}_{\textbf{p}\textbf{q}} = \delta_{\textbf{k}, \textbf{q}-\textbf{p}}\Big\{u_{\textbf{k}}\Big[V_{\text{tot}}(\textbf{p})(u_{\textbf{p}}u_{\textbf{q}} + v_{\textbf{p}}u_{\textbf{q}})+ V_{\text{tot}}(\textbf{q})(v_{\textbf{p}}u_{\textbf{q}} +v_{\textbf{p}}v_{\textbf{q}})\Big] + v_{\textbf{k}}\Big[V_{\text{tot}}(\textbf{q})(u_{\textbf{p}}u_{\textbf{q}} + v_{\textbf{q}}u_{\textbf{p}})  + V_{\text{tot}}(\textbf{p})(v_{\textbf{p}}v_{\textbf{q}} + \\\nonumber u_{\textbf{p}}v_{\textbf{q}})\Big] + (u_{\textbf{k}} +v_{\textbf{k}})V_{\text{tot}}(\textbf{k})(u_{\textbf{p}}u_{\textbf{q}} + v_{\textbf{p}}v_{\textbf{q}})\Big\}.
\end{eqnarray}
\end{widetext}
We denote by $\textbf{p}, \textbf{q}$, the momenta of the outgoing quasi-particles, which are determined by the constraints of energy-momentum conservation, enforced by the delta functions in Eqns.~\ref{landauh}, \ref{aijh}. Here $n^{\text{th}}_{\textbf{k}} =  \frac{1}{e^{\beta E_{\textbf{k}}} -1}$ is the Bose occupation factor for a mode with energy $E_{\textbf{k}}$.


\subsection{Quasi-$2$D dipolar gas}

A quasi-$2$D dipolar gas is created experimentally by employing tight confinement along the axial ($z$) direction, $U(\textbf{r}) =\frac{1}{2}m(\omega^{2}_{x}x^2 + \omega^{2}_{y}y^2+\omega^{2}_{z}z^2)$, where $\omega_{z} \gg \omega_{x}, \omega_{y}$. Here we assume that the confining direction is the same as the direction in which the dipoles are polarized, although our results can be readily generalized to include an arbitrary polarization angle. In the limit of tight confinement, $\mu \ll \hbar\omega_{z}$, with no loss of generality, the density can be expressed as $n(\textbf{r})=n^{\text{2D}}(\rho)\chi(z) = \frac{1}{\sqrt{\pi l^{2}_{z}}}n^{\text{2D}}(\rho)e^{-z^{2}/l^{2}_{z}}$, where $\rho$ and $z$ are the radial and axial coordinates respectively, and $l_{z}$ is a length-scale on the order of the harmonic oscillator wave-length in the axial direction $l_{z} \sim \sqrt{\hbar/m \omega_{z}}$ \cite{fischerdipole}. Integrating out the $z$-direction, one obtains an effective quasi-two dimensional description, which depends on $\rho$ alone. For the homogeneous case we consider here ($\omega_{x} = \omega_{y} = 0$), the Fourier transform of the resulting quasi-$2$D interaction potential reads \cite{fischerdipole, wilson}:
\begin{equation}\label{q2dpot}
V^{\text{q2D}}(k) = g^{\text{q2D}} + g^{\text{q2D}}_{d}F\Big(\frac{{k}l_{z}}{\sqrt{2}}\Big)
\end{equation}
where we define a quasi two-dimensional interaction parameter $g^{\text{q2D}} = \frac{g}{\sqrt{2\pi}l_{z}}$, $k = \sqrt{k^{2}_{x}+k^{2}_{y}}$ is the radial momentum, $g^{\text{q2D}}_{d} = \frac{8\pi}{3\sqrt{2\pi}l_{z}}d^{2}$, and $F(x) = 1 - \frac{3}{2}\sqrt{\pi}x~\text{Erfc}(x)e^{x^{2}}$, where $\text{Erfc}(x)$ is the complimentary error function. When the dipoles are aligned parallel to one another,  the dipole-dipole interaction only depends on the magnitude of the radial momentum. 

The Landau damping rate can be computed in quasi-$2$D simply by replacing $V_{\text{tot}}(\textbf{k})$ with $V^{\text{q2D}}(k)$ and $n_{0}$ with $n^{\text{2D}}_{0}$ in Eqs.~\ref{landauh} and \ref{aijh}. The integrals are now performed over two-dimensional $k$-space.  The Bogoliubov eigenfrequencies and coefficients are: $E_{k} = \sqrt{\epsilon_{k}(\epsilon_{k} + 2 V^{\text{q2D}}(k)n^{\text{2D}}_{0})}$ and  $u_{k} = \sqrt{\frac{1}{2}\Big(\frac{\epsilon_{k} + V^{\text{q2D}}(k)n^{\text{2D}}_{0}}{E_{k}} + 1\Big)}$ and $v_{k} = -\text{sgn}(V^{\text{q2D}}(k))\sqrt{\frac{1}{2}\Big(\frac{\epsilon_{k} + V^{\text{q2D}}(k)n^{\text{2D}}_{0}}{E_{k}} - 1\Big)}$ respectively. 

As discussed by Fischer \cite{fischerdipole} and Santos \textit{et al.} \cite{santos},  the quasi-$2$D dipolar Bose gas is mechanically stable even in the absence of short-range repulsive interactions ($g^{\text{q2D}}=0$). 
Below, we first consider Landau damping of phonons, and derive an energetic criterion for Landau damping in a purely dipolar gas. We contrast the damping in a purely dipolar gas with analogous results in a gas with contact interactions \cite{liu, chung, giorgini, stringari, szepfalusy, schieve}. We then calculate the Landau damping rate for a gas where both contact and dipolar interactions are present, as is the case for current experiments on dipolar $^{52}$Cr, $^{162}$Dy, $^{164}$Dy and $^{168}$Er \cite{pfau2, dysprosium2, dysprosium, erbium}. Finally, we discuss in detail the nature of Landau damping for strong dipolar interactions, where the low energy dispersion acquires a roton-maxon character. 

\section{Landau damping of phonons}

\subsection{Criterion for Landau damping}

We first consider the Landau damping of a phonon with wave-vector $\textbf{k}$ into two quasi-particles with wave-vectors $\textbf{p}$ and $\textbf{q}$.
As shown by Eq.~\ref{landauh}, Landau damping is only allowed provided the conditions for energy and momentum conservation can be simultaneously satisfied. At low energies, the dispersion of the incoming quasi-particle takes the form $E_{k} \approx \hbar ck$, where $c$ is the phonon velocity in the gas, which will be defined subsequently. 

Momentum conservation (enforced by the delta function in Eq.~\ref{aijh}) implies that $\textbf{q} = \textbf{k}+\textbf{p}$, which can be expanded as $q = |\textbf{q}| \approx p + k \cos(\theta)$ in the long wave-length limit $\hbar k \ll mc$. Here $\theta$ is the angle between the vectors $\textbf{p}$ with $\textbf{k}$. Likewise, $E_{\textbf{q}}-E_{\textbf{p}} \approx  \hbar v_{g}(p)k\cos(\theta)$ where $v_{g}(p) = 1/\hbar~\partial E_{p}/\partial p$ is the group velocity of the excitations \cite{stringari}.  Energetically, Landau damping can occur only if
\begin{equation}\label{econs}
\frac{c}{v_{g}(p)} \leq 1,
\end{equation}
\textit{i.e} the velocity of the incoming quasi-particle has to be less than the group velocity of the decaying excitations. We note that this criterion is not limited to damping of phonons in Bose condensates; rather it is completely general, and also applies to Landau damping in Fermi liquids or plasmas \cite{pines}.

At large values of $\hbar p \gg mc$, the dispersion for both dipolar and contact interactions becomes free-particle-like, $E_{p} \rightarrow \hbar^{2}p^{2}/2m$ and thus $v_{g}(p) \rightarrow \hbar p/m$. As $p \rightarrow \infty$, Eq.~\ref{econs} is always satisfied, and Landau damping can occur. 

However, at small quasi-particle wave-vectors, the dispersion of the dipolar gas is dramatically different from a gas with contact interactions. For a gas with purely contact interactions, the dispersion takes the form: $E_{p}/\hbar \approx c p + \frac{\hbar^{2}p^{3}}{8m^{2}c}$, where $c = \sqrt{g^{\text{q2D}}n^{\text{2D}}_{0}/m}$ is the phonon velocity. Thus the criterion for energy conservation as $p/mc \rightarrow 0$ reads: $c/v_{g} \sim 1 - {\cal{O}}(p^{2}) < 1$ and Landau damping is allowed. By contrast, the dipolar dispersion $E_{p}$ is given by:
\begin{equation}\label{dipexp}
E_{p}/\hbar \approx c_{d} p - \frac{3}{4}\sqrt{\frac{\pi}{2}}l_{z}c_{d}p^{2}  + \frac{\hbar^{2}p^{3}}{8m^{2}c_{d}}
\end{equation}
where we have introduced a speed of sound associated with the dipolar interaction $c_{d} = \sqrt{g^{\text{q2D}}_{d}n^{\text{2D}}_{0}/m}$. As $p \rightarrow 0$, $c_{d}/v_{g} \sim 1 + {\cal{O}}(p) > 1$, and Landau damping is thus forbidden by energy-momentum conservation. 

Landau damping becomes possible when the cubic term in the expansion of Eq.~\ref{dipexp} dominates over the quadratic term. 
Equating the quadratic and cubic terms in Eq.~\ref{dipexp}, we find that Landau damping occurs only when $p$ exceeds a critical value:
\begin{equation}\label{pcrit}
p > p_{\text{crit}} = \frac{2\sqrt{2\pi}l_{z}m^{2}c^{2}_{d}}{\hbar^{2}} = \frac{2}{l_{z}}\frac{E_{\text{int}}}{E_{\text{ho}}}
\end{equation}
where $E_{\text{int}} = g^{\text{q2D}}_{d}n^{\text{2D}}_{0}$ is the mean-field interaction energy and $E_{\text{ho}} = \hbar^{2}/ml^{2}_{z} = \hbar\omega_{z}$ is the axial harmonic oscillator energy. In other words, a phonon with wave-vector $\bf{k}$ can only Landau damp into excitations with wave-vectors $\bf{p}$ and $\bf{p}+\bf{k}$ for $p > p_{\text{crit}}$. 

\subsection{Temperature dependence of damping rate in a purely dipolar gas}

To obtain the temperature dependence of the phonon damping rate, we expand Eq.~\ref{aijh} for small values of $k$. Introducing dimensionless quantities for the interaction strength, momentum and temperature, $\tilde c_{d} = \sqrt{E_{\text{int}}/E_{\text{ho}}}$, $\tilde k = kl_{z}/\sqrt{2}$, and $\tau = k_{B}T/E_{\text{ho}}$ respectively, the Landau damping rate for a phonon with momentum $k$ and energy $E_{k} = \hbar c_{d} k$ in a quasi-$2$D dipolar gas reads:

\begin{equation}\label{landaufinal}
\frac{\hbar\Gamma_{L}}{E_{k}} = \frac{\tilde c_{d}}{4\pi n^{\text{2D}}_{0}l^{2}_{z}}{\cal{I}}(\tau)
\end{equation}
\\
where the integral ${\cal{I}}(\tau)$ is given by:

\begin{widetext}
\begin{eqnarray}\label{landauint}
{\cal{I}}(\tau) = \frac{1}{\tau}\int^{\infty}_{\sqrt{2}\tilde c^{2}_{d}} dx~ x \cot(\theta_{c}) \frac{1}{(e^{u/2\tau} -e^{-u/2\tau})^{2}}\times\Big(\frac{x^{2}+\tilde c^{2}_{d}F(x)}{u}(1+F(x)) - \frac{\tilde c^{2}_{d}}{u}F^{2}(x)\text{sgn}(F(x)) +\\\nonumber \frac{\tilde c^{4}_{d}F(x)}{v(x)\text{sgn}(F(x))}\Big[\frac{-F(x)v(x) + u\partial_{x}F(x)}{u\sqrt{u^{2} + \tilde c^{4}_{d}F(x)^{2}}}\Big]\Big)^{2}
\end{eqnarray}
\end{widetext}
where $F(x)$ is defined in Eq.~\ref{q2dpot}, $u = \sqrt{x^2(x^2+ 2 \tilde c^{2}_{d}F(x))}$, $v(x) = \partial u/\partial x$ and $\theta_{c} = \sqrt{2}\tilde c_{d}/v(x)$. For a gas with purely contact interactions, we simply set $F(x) = 1$, and $\tilde c_{d} \rightarrow \tilde c$ where $\tilde c$ is similarly defined, and the integral over $x$ runs from $0$ to $\infty$. 

\begin{figure}
\begin{picture}(200, 120)
\put(10, -10){\includegraphics[scale=0.5]{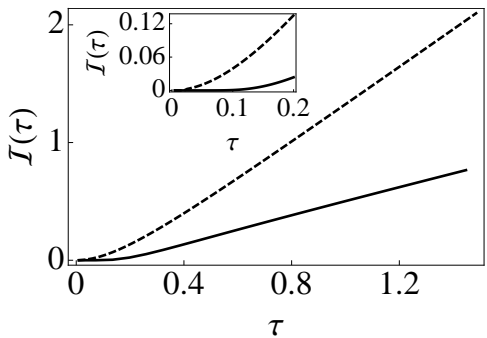}}
\end{picture}
\caption{\label{dampingrate} The integral ${\cal{I}}(\tau)$ (Eq.~\ref{landauint}) plotted as a function of the reduced temperature $\tau = k_{B}T/E_{\text{ho}}$ in quasi-$2$D. The solid line is for a purely dipolar gas while the dashed line is for a gas with contact interactions. We choose the dimensionless speed of sound for the dipolar and contact interactions $\tilde c_{d} = \tilde c = 1$ to enable comparison between the two cases. At high temperatures the damping rates scales linearly with $T$ in both cases, with the dipolar damping rate being consistently smaller than that for a gas with contact interactions. Inset is a zoom-in at low temperatures showing the dramatic suppression of the dipolar damping rate compared to that in a gas with contact interactions.} 
\end{figure}

In Fig.~\ref{dampingrate} we plot the integral ${\cal{I}}(\tau)$ obtained by numerically integrating Eq.~\ref{landauint} at fixed $\tilde c_{d}$. The corresponding damping rate for a gas with purely contact interactions ($\tilde c= 1$) is also shown for comparison. The phonon damping rate in a purely dipolar gas remains consistently lower than that for a gas with contact interactions at all temperatures. At high temperature ($\tau > 1$), the damping rates appear to scale linearly with temperature in both cases, while at low temperature ($\tau \ll 1$), the damping rate in a dipolar gas drops significantly faster than the damping rate for a gas with contact interactions. In particular, the dipolar damping rate is found to scale as $e^{-1/\tau}$ for $\tau \ll 1$. This is in stark contrast with the corresponding damping rate in a gas with purely contact interactions, which scales as $T^{4}$ in $3$D \cite{liu, giorgini, stringari,szepfalusy} and $T^{2}$ in $2$D \cite{chung}. 

As discussed previously, for a dipolar gas, modes with energy less than $E_{\text{crit}} = \hbar^{2}p^{2}_{\text{crit}}/2m$ do not contribute to Landau damping. At low temperatures $k_{B}T \ll E_{\text{crit}}$, the occupation of modes with energy $E > E_{\text{crit}}$  scales as  $n^{\text{th}}_{k > k_{\text{crit}}} \sim e^{-\beta E_{k}}$, and subsequently the damping is suppressed. As shown in the inset of Fig.~\ref{dampingrate}, the low temperature damping rate in a purely dipolar gas grows much more slowly than the $T^{2}$ scaling for a gas with contact interactions \cite{chung}. Consequently, phonons in a purely dipolar gas are virtually undamped at low temperatures.

At high temperatures, $k_{B}T  \gg E_{\text{crit}}$, the damping rate scales linearly with temperature as in the purely contact case, however the magnitude of the damping rate at fixed $\tilde c_{d}$ is smaller than that of a gas with the purely contact interactions, as phonons cannot damp into excitations with momentum less than $p_{\text{crit}}$.

\subsection{Landau damping with contact and dipolar interactions}

\begin{figure}
\begin{picture}(150, 200)
\put(-20, 100){\includegraphics[scale=0.41]{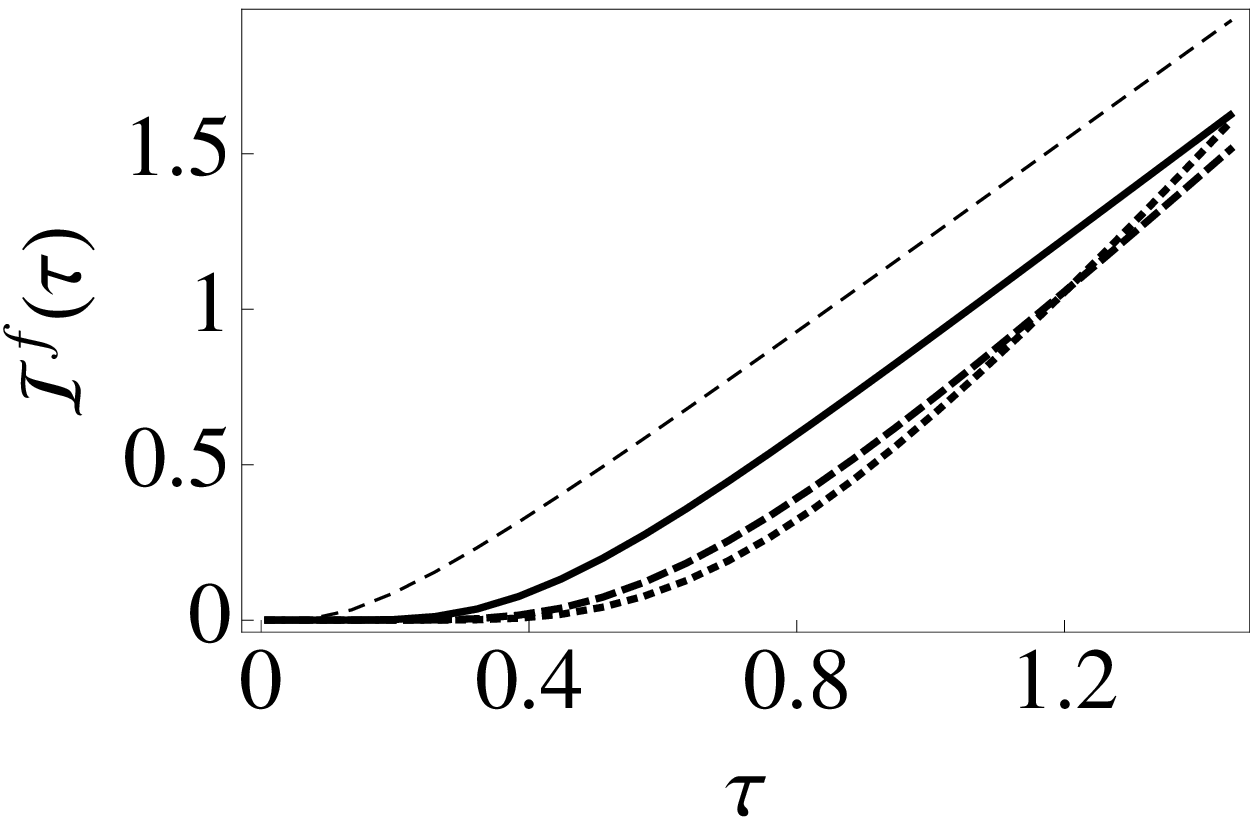}}
\put(-10, -10){\includegraphics[scale=0.39]{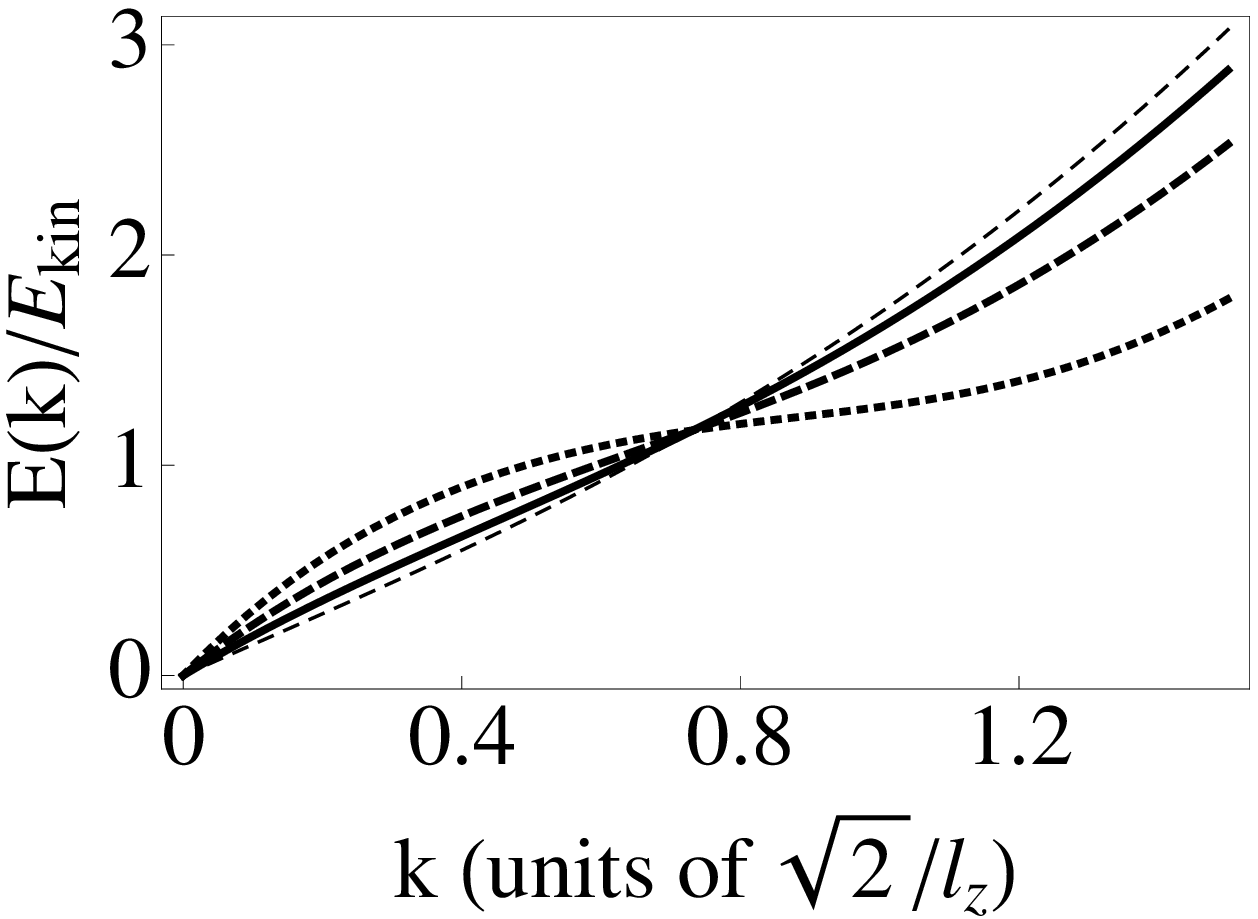}}
\end{picture}
\caption{\label{dampingrate2} (Top Plot) Temperature dependence of the phonon damping rate plotted versus reduced temperature $\tau = k_{B}T/E_{\text{ho}}$ in a quasi-$2$D gas with dipolar and contact interactions present. The contact interaction is held fixed ($\tilde c = 1$) in all the curves and the dimensionless ratio $\tilde g = g^{\text{q2D}}_{d}/g^{\text{q2D}}$ is varied. From top to bottom: (thin dashed) $\tilde g = 0.1$, (thick, solid) $\tilde g = 1$, (thick, dashed) $\tilde g  = 2.5$, (thick dotted) $\tilde g = 5$. (Bottom Plot) The energy momentum dispersion corresponding to the same values of $\tilde g$ and $\tilde c$ as in the top figure.} 
\end{figure}

In Fig.~\ref{dampingrate2} we show the temperature dependence of the Landau damping rate for phonons in a gas with contact and dipolar interactions. Once again the damping rate can be expressed as $\frac{\hbar\Gamma_{L}}{E_{k}} = \frac{\tilde c}{4\pi n^{\text{2D}}_{0}l^{2}_{z}}{\cal{I}}^{f}(\tau)$, where $\tilde c = \sqrt{g^{\text{q2D}}n^{\text{2D}}_{0}/E_{\text{ho}}}$ parameterizes the contact part of the interaction, $E_{k} = \hbar c k$, and the integrand ${\cal{I}}^{f}(\tau)$ contains the temperature dependence. The full expression for ${\cal{I}}^{f}(\tau)$ is rather cumbersome, and is not shown here.  We also introduce a dimensionless ratio $\tilde g = g^{\text{q2D}}_{d}/g^{\text{q2D}}$, which parametrizes the relative strength of the dipole-dipole and the contact interaction.

In Fig.~\ref{dampingrate2}, $\tilde c$ is held fixed and we vary $\tilde g$. The corresponding energy-momentum dispersion, normalized to $E_{\text{ho}}$, is shown on the bottom. Somewhat counterintuitively, we find that at fixed temperature, increasing the dipolar interaction  \textit{decreases} the phonon damping rate. As the  dipolar interaction strength is increased, $p_{\text{crit}}$ increases, and the number of available modes for Landau damping decreases. When the interactions become large enough such that $E_{\text{crit}} = \hbar^{2}p^{2}_{\text{crit}}/2m \gg k_{B}T$, the damping rate is strongly suppressed as the available modes for Landau damping are  largely unoccupied. For strong dipole-dipole interactions, phonons are virtually undamped at low temperatures. At higher temperatures $\tau \sim 1$, the damping rate again scales linearly with $T$. 

Before discussing the damping of the roton-maxon excitations, we briefly comment on the validity of the results presented in this section. Strictly speaking, the quasi-$2$D ansatz we employ here is only valid provided that $\eta = m (g^{\text{q2D}} + g^{\text{q2D}}_{d})/4\pi\hbar^{2} \ll 1$. As the interaction strength is increased, the Gaussian ansatz for the axial wave-function needs to be modified, for example by replacing the width of the Gaussian $l_{z}$ with a variational parameter that depends on the interaction strength \cite{petrov}. While this will quantitatively change our results, it will not affect the qualitative physics. We also expect our results to hold as long as the temperature $T \leq \mu$. At higher temperatures and strong interactions, the condensate becomes significantly depleted, our starting point of treating the coupling between the condensed and non-condensate atoms as a weak perturbation to Bogoliubov theory is no longer valid. Moreover, exchange interactions between non-condensed atoms alter the Bogoliubov equations of motion. A systematic understanding of how these effects  modify the physics of the dipolar gas is currently being developed \cite{baillie}.

\section{Landau damping in the maxon-roton regime}

Thus far we have focused on the damping of low energy phonons in a quasi-$2$D dipolar gas. One of the novel features of a dipolar gas is that for sufficiently strong dipole-dipole interactions, the dispersion develops a roton-maxon character \cite{santos, fischerdipole, dell} at intermediate momenta. We now discuss the Landau damping of these excitations.
\begin{figure}
\begin{picture}(150, 200)
\put(-20, 100){\includegraphics[scale=0.41]{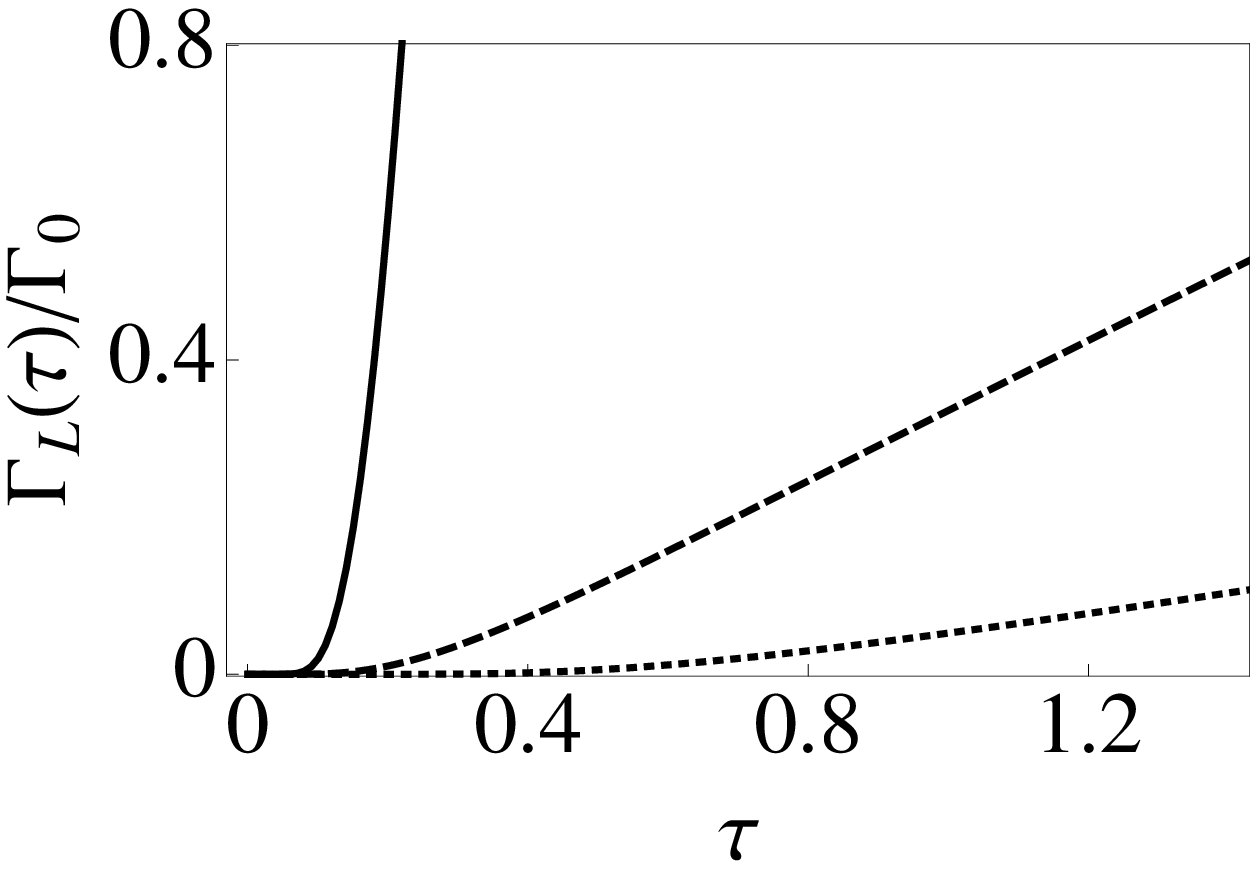}}
\put(-20, -10){\includegraphics[scale=0.39]{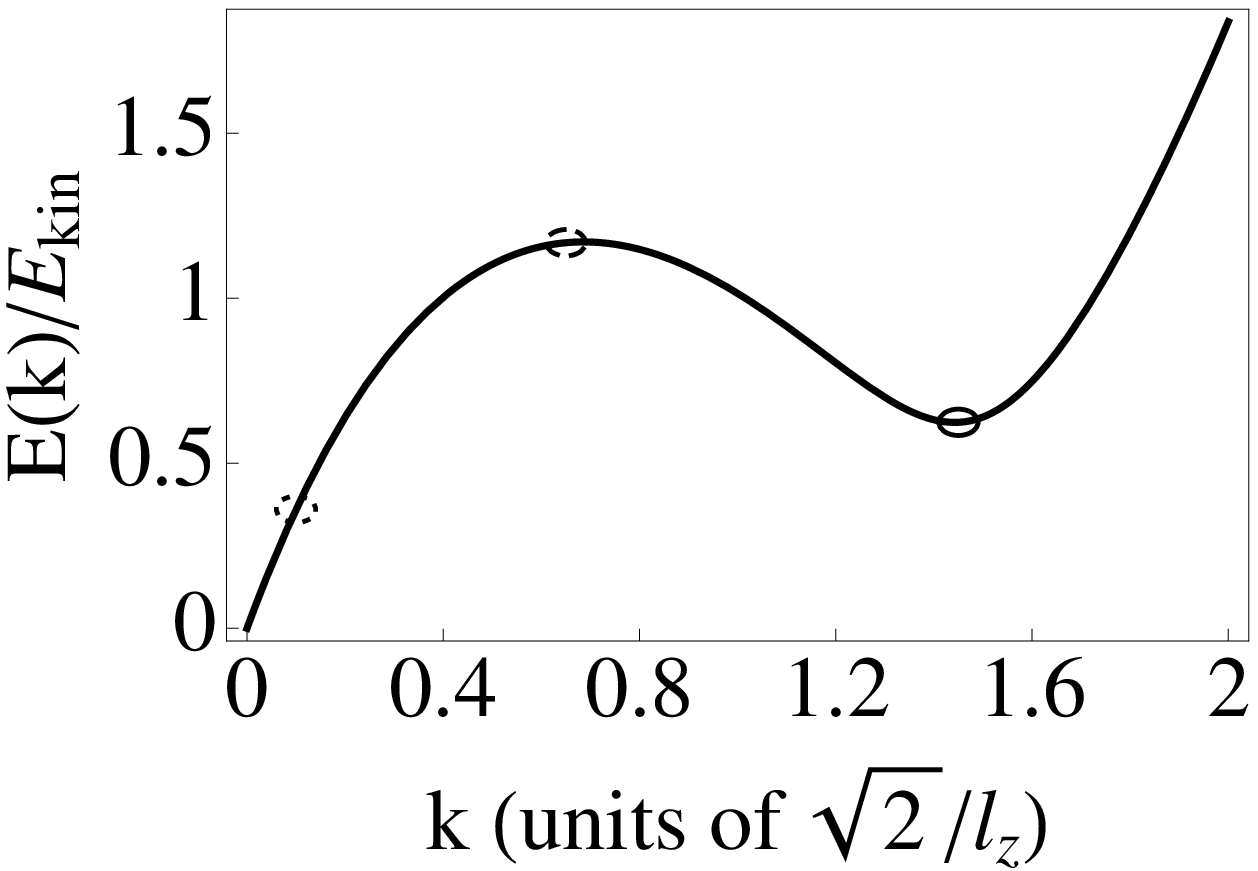}}
\end{picture}
\caption{\label{dampingrateroton} (Top Plot) Temperature dependence of the Landau damping rate at three different momenta corresponding to the phonon (dotted), maxon (dashed) and roton (solid) for a quasi-$2$D gas with contact and dipolar interactions.  The damping rate is normalized to $\hbar\Gamma_{0} = \sqrt{2/\pi}E_{\text{ho}}\tilde c^{2} (a/l_{z})$. The contact interaction pararmetrized by $\tilde c = \sqrt{g^{\text{q2D}}n^{\text{2D}}_{0}/E_{\text{ho}}} =1$ and the dimensionless ratio $\tilde g = g^{\text{q2D}}_{d}/g^{\text{q2D}} = 7.2$. (Bottom Plot) The dispersion for $\tilde c = 1$ and $\tilde g = 7.2$ plotted versus $k$. The solid, dashed and dotted circles correspond to the values of the momenta used in the Top plot.}
\end{figure}

In Fig.~\ref{dampingrateroton} we plot the Landau damping rate at intermediate momenta for strong dipole-dipole interactions. The damping rate is normalized to $\hbar\Gamma_{0} = \sqrt{2/\pi}E_{\text{ho}}\tilde c^{2} (a/l_{z})$. The curves are obtained by numerically integrating Eqn.~\ref{landauh} using Eq.~\ref{aijh}. As before, we fix the contact part of the interaction ($\tilde c =1$) and vary the ratio $\tilde g$. The corresponding phonon damping rate for the same value of $\tilde g$ is shown for comparison. At the value of $\tilde g$ chosen here, the critical momentum at which Landau damping of phonons is allowed is $p_{\text{crit}} \sim 2 \hbar\sqrt{2}/l_{z}$ (Eq.~\ref{pcrit}). Phonon damping is therefore highly suppressed at strong dipole-dipole interactions as virtually none of the modes that are available for Landau damping are thermally occupied.

To understand the damping rate at intermediate momenta, first we show that rotons cannot scatter off of phonons. To see this, consider an incoming roton with momentum $\textbf{k}$, scattering off of a phonon with momentum $\textbf{p}$. The criterion for Landau damping now reads $c/v^{\text{r}}_{g}(k) \leq 1$ where $v^{\text{r}}_{g}(k)$ is the group velocity of the rotons. Near the roton minimum, the excitations are free-particle like with a small gap, and the group velocity scales as $|k - k_{r}|$ where $k_{r}$ corresponds to the momentum at the roton minimum. As $k \rightarrow k_{r}$, $v^{\text{r}}_{g} \rightarrow 0$, and Landau damping is thus forbidden. 

Solving the energy conservation criterion numerically, we find that Landau damping of rotons turns on at intermediate momenta, $p^{\text{r}}_{\text{crit}} \geq 0.5 \sqrt{2}\hbar/l_{z}$, which is consistent with the location of the maxon peak in the excitation spectrum. This critical momentum is considerably smaller than $p_{\text{crit}}$ for phonon damping, and as a result, at any given temperature, a larger number of modes contribute to the Landau damping of rotons. Hence, the roton damping rate is significantly higher than the damping rate for phonons. The damping rate for the maxon mode lies between the damping rates for the phonon and roton modes.

\begin{figure}
\begin{picture}(150, 100)
\put(-20, -10){\includegraphics[scale=0.4]{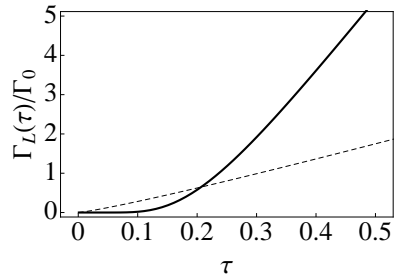}}
\end{picture}
\caption{\label{rotondampcompare} Temperature dependence of the Landau damping rate at $\tilde g = 7.2$ and $k = 1.45$ (corresponding to the roton minimum on the plot in Fig.~\ref{dampingrateroton} (normalized to $\hbar\Gamma_{0} = \sqrt{2/\pi}E_{\text{ho}}\tilde c^{2} (a/l_{z})$) for a quasi-$2$D gas with contact and dipolar interactions. The dashed curve shows the damping rate for the same value of momentum, but with no dipolar interactions $\tilde g = 0$. For our parameters, the roton instability (where the roton minimum is at zero energy) occurs at $\tilde g = 7.5$.}
\end{figure}

Finally, in Fig.~\ref{rotondampcompare}, we compare the damping rate at fixed momentum for a gas with strong and weak dipole-dipole interactions. Recall that in the long wave-length limit, we found that the damping rate \textit{decreases} with increasing dipole-dipole interactions over the temperature range $0 < \tau \lesssim 1$ (see Fig.~\ref{dampingrate2}). By contrast, here we find that the damping rate at intermediate momenta \textit{increases} with increasing dipolar interactions, except at very low temperatures. At very low temperatures, modes with energies $\sim E_{p^{r}_{\text{crit}}}$ are unoccupied, and the rotons are undamped. At intermediate temperatures, the damping rate grows linearly with $T$ but the slope is much larger for the rotons  (solid curve) than for the excitations in a non-dipolar gas at the same momentum (dashed curve). 

The softening of the roton mode $E_{k_{\text{r}}} \rightarrow 0$ is associated with a divergence of the Bogoliubov amplitudes $u_{k}$ and $v_{k}$. As the damping rate is directly proportional to the Bogoliubov amplitudes, via the matrix elements $A_{\textbf{k}\textbf{q}}$ in Eq.~\ref{aijh}, the damping rate also increases near the roton minimum for strong dipole-dipole interactions. We note however that our approach breaks down well before the roton instability is reached, as the diverging Bogoliubov amplitudes imply significant condensate depletion, and our assumption of $n_{\text{tot}} \sim \phi^{2}_{0}$ becomes invalid. 

Boudgemaa and Shlyapnikov \cite{boudgemaa} argue that for low temperatures $T \ll \Delta_{\text{r}}$, where $\Delta_{\text{r}}$ is the roton gap, the Bogoliubov approach is valid as long as $\eta \ll 1$. At higher temperatures or stronger interactions where the roton gap is small $T \gg \Delta_{\text{r}}$, thermal effects begin to dominate. For the interaction parameters of Fig.~\ref{dampingrateroton}, the roton gap $\Delta_{\text{r}} \sim 0.65 E_{\text{kin}}$ (see bottom panel in Fig.~\ref{dampingrate2}). Somewhat surprisingly, the damping rate for the roton mode is large even at temperatures smaller than $\Delta_{\text{r}}$ suggesting that roton excitations may be short-lived at finite temperatures. 

\section{Experimental significance}

We now briefly discuss the possibility of observing the physics described above in experiments. Experimentalists can create a single $2$D pancake trap by confining atoms in an optical dipole trap with frequencies $\omega_{r} \sim 2\pi \times 10$ Hz, and $\omega_{z} \sim 2\pi \times 10^{3}$~Hz \cite{chengstruc}. For bosonic $^{164}$Dy, which was recently Bose-condensed by Lu and co-workers \cite{dysprosium2}, this yields a transverse confinement length $l_{z} = \sqrt{\hbar/m\omega_{z}} = 0.1\mu$m. Experiments typically operate at temperatures $T \sim 10$nK, which yields a $\tau = k_{B}T/E_{\text{ho}} \sim 0.1$, and corresponding $2$D densities of $n^{\text{2D}}_{0} \sim 5 \times 10^{14}~\text{cm}^{-2}$ \cite{chengstruc}. This implies that $\tilde c = \sqrt{g^{\text{q2D}}n^{\text{2D}}_{0}/E_{\text{ho}}} \sim 1$ can be achieved for scattering lengths $a/l_{z} \sim 0.02 \ll 1$ \cite{boudgemaa}. Choosing a typical scattering length of $a \sim 100 a_{B}$, $\hbar\Gamma_{0} \sim  1-10$Hz.

Damping of excitations in the non-dipolar gas was observed by Katz \textit{et al.} \cite{katz} who used Bragg spectroscopy to study the momentum dependence of the Beliaev damping rate of phonons. By tuning the energy difference and the angle between the Bragg beams, experimentalists excited quasi-particles at different energies and momenta. After a short time-of-flight, a halo of scattered atoms was observed at momenta intermediate between the condensate and the momentum of the atoms excited initially. By measuring the ratio of the number of scattered atoms to the initial cloud of atoms excited by the probe beams, as a function of energy and momentum, Katz and co-workers were able to extract the damping rate \cite{katz}. A similar technique may be used to detect Landau damping in the dipolar gas.

The energy resolution achieved in the experiment was $\sim $~kHz, and the momentum resolution was $\sim 2\pi/\zeta$ where $\zeta \sim 0.25\mu$m is the coherence length of the Bose condensate. The roton mode is located at $kl_{z} \sim 1.5$, and hence lies above the typical momentum resolution in experiments. In addition to spectroscopic probes, Landau damping also influences the decay of collective excitations in trapped gases \cite{jin2, ketterle}. 

\section{Summary and Conclusions}

In summary, we have developed a theory of damping of low energy excitations in a Bose gas interacting with long range interactions, generalizing previous works on Bose gases interacting with short range forces \cite{chung, giorgini}. We work in the collisionless limit, where the damping is provided by scattering of a quasi-particle with other excitations, mediated by the Bose condensate.  Following the time-dependent mean-field approach of Giorgini \cite{giorgini}, we derive a coupled system of equations describing the dynamics of the condensed and non-condensed atoms, within the Popov approximation. We then solve these equations perturbatively in the interaction strength, to obtain the expression for the Landau damping rate in a Bose gas with long range interactions. 

Focussing on the homogeneous gas and neglecting exchange interactions between non-condensed atoms, we find that the nature of the low energy quasi-$2$D dipolar dispersion forbids a phonon from decaying into modes with momenta lower than a certain threshold. This leads to a dramatic suppression of Landau damping at low temperatures. For stronger dipole-dipole interactions, we find that phonons are virtually undamped over a large temperature regime. By contrast, at intermediate momenta and strong interactions, where the dispersion develops a roton-maxon feature, we find much higher damping rates.

Finally, we remark that experiments on collective modes in trapped gases typically also measure the temperature dependence of the frequency shift of the collective modes, in addition to the damping rates \cite{jin2, ketterle}. These frequency shifts are related to many body effects which go beyond the Popov approximation \cite{giorgini2, fedichev, shenoy, griffinbook, griffin}. A systematic study of these effects and how they affect the frequencies of the collective excitations in the dipolar gas has not yet been performed, and is a promising direction for future study \cite{blakienew}.  

\section{Acknowledgements}

It is a pleasure to thank  Benjamin Lev, Kristian Baumann and Mingwu Lu for several discussions regarding the experimental implications of the physics described above. We are also grateful to Sankar Das Sarma for his careful reading of the manuscript and for suggesting improvements. S.N. is supported by JQI-NSF-PFC, AFOSR-MURI, and ARO-MURI. R. M. W. acknowledges support from an NRC postdoctoral Fellowship.


\begin{thebibliography}{99}
\bibitem{pfau2} T. Koch, T. Lahaye, J. Metz, B. Fr\"ohlich, A. Griesmaier and T. Pfau, Nature Physics \textbf{4} 218 (2008).
\bibitem{dysprosium2} M. Lu, N. Q. Burdick, S. H. Youn and B. L. Lev, Phys. Rev. Lett. \textbf{107} 190401 (2011). 
\bibitem{dysprosium} M. Lu, N.Q. Burdick and B. L. Lev, Phys. Rev. Lett. \textbf{108} 215301 (2012).
\bibitem{erbium} K. Aikawa, A. Frisch, M. Mark, S. Baier, A. Rietzler, R. Grimm and F. Ferlaino, Phys. Rev. Lett. \textbf{108} 210401 (2012).
\bibitem{lahayereview} T. Lahaye, C. Menotti, L. Santos, M. Lewenstein and T. Pfau, Rep. Prof. Phys. \textbf{72} 126401 (2009).
\bibitem{pfau} T. Lahaye, J. Metz, B. Fro\"hlich, T. Koch, M. Meister, A. Griesmaier, T. Pfau, H. Saito, Y. Kawaguchi and M. Ueda ,Phys. Rev. Lett. \textbf{101}, 080401 
(2008).
\bibitem{wilson} C. Ticknor, R. M. Wilson and J. L. Bohn Phys.Rev. Lett. \textbf{106} 065301 (2011).
\bibitem{abad} M. Abad, M. Guilleumas, R. Mayol, M. Pi and D. M. Jezek, Phys. Rev. A \textbf{79} 063622 (2009).
\bibitem{cooper} S. Komineas and N. R. Cooper, Phys. Rev. A \textbf{75} 023623 (2007).
\bibitem{santos} L. Santos, G. V. Shlyapnikov and M. Lewenstein, Phys. Rev. Lett. \textbf{90} 250403 (2003).
\bibitem{giovanazzi} S. Giovanazzi and D. H. J. O'Dell, Eur. Phys. D \textbf{31} 439 (2004).
\bibitem{mazzanti} F. Mazzanti, R. E. Zillich, G. E. Astrakharchik and J. Boronat, Phys. Rev. Lett. \textbf{102} 110405 (2009).
\bibitem{marisreview} H. J. Maris Rev. Mod. Phys. \textbf{49} 341(1977). 
\bibitem{mills} N. G. Mills, R. A .Sherlock and A. F. G. Wyatt, Phys. Rev. Lett. \textbf{31} 687 (1974).
\bibitem{griffinbook} A. Griffin, \textit{Excitations in a Bose-condensed Liquid} (Cambridge University Press, New York, 1993).
\bibitem{Beliaev} S. T. Belieav Sov. Phys. JETP \textbf{34} 323 (1958).
\bibitem{martin} P. C. Hohenberg and P. C. Martin, Ann. Phys. (NY) \textbf{34} 291 (1965).
\bibitem{szepfalusy} P. Sz\'epfalusy and I. Kondor, Ann. Phys. \textbf{82} 1-53 (1974).
\bibitem{fedichev} P. O. Fedichev and G. V. Shlyapnikov, Phys. Rev. A \textbf{58} 3146 (1998).
\bibitem{giorgini} S. Giorgini, Phys. Rev. A \textbf{57} 2949 (1998).
\bibitem{giorgini2} S. Giorgini, Phys. Rev. A \textbf{61} 063615 (2000).
\bibitem{schieve} W. V. Liu and W. C. Schieve, eprint. cond-mat/9702122.
\bibitem{chung} M-C. Chung and A. B Bhattacherjee, New. J Phys \textbf{11} 123012 (2009).
\bibitem{shlyapnikov} P. O. Fedichev, G. V. Shlyapnikov and J. T. M. Walraven, Phys. Rev. Lett. \textbf{80} 2269 (1998).
\bibitem{stringari} L. P. Pitaevskii and S. Stringari, Phys. Lett. A \textbf{235} (1997).
\bibitem{liu} W. V. Liu, Phys. Rev. Lett. \textbf{79} 4056 (1997). 
\bibitem{griffin} E. Zaremba, A. Griffin and T. Nikuni, Phys. Rev. A \textbf{57} 4695 (1998).
\bibitem{shenoy} V. B. Shenoy and T. -L. Ho, Phys. Rev. Lett. \textbf{80} 3895 (1998).
\bibitem{stoof} M. J. Bijlsma and H. T. C. Stoof, Phys. Rev. A \textbf{60} 3973 (1999).
\bibitem{pethick} \textit{Bose-Einstein Condensation in Dilute Gases} C. J. Pethick and H. Smith, Cambridge University Press, 2002.
\bibitem{popov} V. N. Popov, Theoretical and Mathematical Physics \textbf{11} 478 (1972).
\bibitem{griffin3} A. Griffin, T. Nikuni and E. Zaremba \textit{Bose-Condensed Gases at Finite Temperatures} (Cambridge University Press, Cambridge, UK, 2009).
\bibitem{jin} D. S. Jin, J. R. Ensher, M. R. Matthews, C. E. Wieman, and E. A. Cornell, Phys. Rev. Lett. \textbf{77} 420 (1996).
\bibitem{stringari1} S. Stringari, Phys. Rev. Lett. \textbf{77} 2360 (1996).
\bibitem{ketterle} R. Onofrio, D. S. Durfee, C. Raman, M. Kohl, C. E. Kuklewicz, and W. Ketterle, Phys. Rev. Lett. \textbf{84} 810 (2000). 
\bibitem{ketterle2} M.-O. Mewes, M. R. Andrews, N. J. van Druten, D. M. Kurn, D. S. Durfee, C. G. Townsend and W. Ketterle, Phys. Rev. Lett. \textbf{77} 988 (1996).
\bibitem{jin2} D. S. Jin, M. R. Matthews, J. R. Ensher, C. E. Wieman and E. A Cornell, Phys. Rev. Lett. \textbf{78} 764 (1997).
\bibitem{blakie} P. B. Blakie, D. Baillie and R. N. Bisset, Phys. Rev. A \textbf{86} 021604 (R) (2012).
\bibitem{fischerdipole} U. R. Fischer, Phys. Rev. A \textbf{73} 031602 (R) (2006).
\bibitem{wilson2} R. M. Wilson, S. Ronen, J. L. Bohn and H. Pu, Phys. Rev. Lett. \textbf{100} 245302 (2008).
\bibitem{natu1} S. S. Natu and S. Das Sarma, Phys. Rev. A \textbf{88} 031604 (R) (2013).
\bibitem{ryan1} R. M. Wilson and S. S. Natu (in preparation).
\bibitem{caveat} By assuming the condensate density is real, we do not consider ground states of the condensate which include vortices, but these can be readily incorporated into our approach. 
\bibitem{ronen} S. Ronen, D. C. E. Bortolotti and J. L. Bohn, Phys. Rev. A \textbf{74} 013623 (2006).
\bibitem{blakie2} D. Baillie and P. B. Blakie, Phys. Rev. A \textbf{82} 033605 (2010).
\bibitem{ticknor} C. Ticknor, Phys. Rev. A \textbf{85} 033629 (2012).
\bibitem{pines} D. Pines and P. Nozi\`eres, \textit{The Theory of Quantum Liquids}, (Westview Press, Boulder, Colorado, 1989).
\bibitem{petrov} D. S. Petrov, M. Holzmann and G. V. Shlyapnikov Phys. Rev. Lett. \textbf{84} 2551 (2000).
\bibitem{baillie} D. Baillie and P. B. Blakie, Phys. Rev. A \textbf{86} 041603 (R) (2012).
\bibitem{blakienew} P. B. Blakie, D. Baillie and R. N. Bisset, Phys. Rev. A \textbf{88} 013638 (2013).
\bibitem{dell} D. H. J. O'Dell, S. Giovanazzi and G. Kurizki, Phys. Rev. Lett. \textbf{90} 110402 (2003).
\bibitem{boudgemaa} A. Boudgemaa and G. V. Shlyapnikov, eprint.arXiv: 1212.1136 (2012).
\bibitem{katz} N. Katz, J. Steinhauer, R. Ozeri and N. Davidson, Phys. Rev. Lett. \textbf{89} 220401 (2002). 
\bibitem{chengstruc} C-L. Hung, X. Zhang, L-C. Ha, S-K. Tung, N. Gemelke and C. Chin, New J. Phys \textbf{13} 075019 (2011).

\end{thebibliography}
\end{document}